\documentclass[times]{aa}
\usepackage{psfig}
\usepackage{natbib}

\bibpunct{(}{)}{;}{a}{}{,}

\begin{document}

\def\eps{\varepsilon}
\def\aap{A\&A}
\def\apj{ApJ}
\def\apjl{ApJL}
\def\mnras{MNRAS}
\def\aj{AJ}
\def\nat{Nature}
\def\aaps{A\&A Supp.}
\def\me{m_\e}
\def\lesssim{\mathrel{\hbox{\rlap{\hbox{\lower4pt\hbox{$\sim$}}}\hbox{$<$}}}}
\def\gtrsim{\mathrel{\hbox{\rlap{\hbox{\lower4pt\hbox{$\sim$}}}\hbox{$>$}}}}
\def\degs{$^\circ $}
\def\msun{{\,M_\odot}}
\def\lsun{{\,L_\odot}}
\def\sgra{Sgr~A$^*$}
\def\simlt{\lower.5ex\hbox{$\; \buildrel < \over \sim \;$}}
\def\simgt{\lower.5ex\hbox{$\; \buildrel > \over \sim \;$}}
\newcommand\schodel{Sch\"odel~}

\def\del#1{{}}

\def\C#1{#1}

\title{Bright stars and an inactive disk in Sgr~A$^*$ and other\\
dormant  galaxy centers. I. The optically thick disk.} 
\titlerunning{Bright stars and an optically thick disk in Sgr~A$^*$}  \author{Jorge Cuadra\inst{1} \and Sergei Nayakshin\inst{1} \and
Rashid Sunyaev\inst{1,2}} \authorrunning{J. Cuadra et al.}
\institute{$^1$Max-Planck-Institut f\"{u}r Astrophysik,
Karl-Schwarzschild-Str.1, D-85741 Garching, Germany\\
$^2$Space Research Institute, Moscow, Russia}
\date{\today} \offprints{jcuadra@MPA-Garching.MPG.DE}

\abstract{Cold inactive disks are believed to exist in Low Luminosity
AGN (LLAGN). They may also exist in the nuclei of inactive galaxies
and in the center of our own Galaxy. These disks would then be
embedded in the observed dense nuclear stellar clusters. Making the
simplest assumption of an optically thick disk, we explore several
ways to detect the disk presence through its interaction with the
cluster.  First of these is the eclipses of close bright stars by the
disk. The second is the increase in the infrared flux of the disk due
to illumination of its surface by such stars during close
passages. Finally the surface brightness of the star cluster should
show an anisotropy that depends on the inclination angle of the
disk. We apply the first two of the methods to \sgra, the super
massive black hole in our Galactic Center. Using the orbital
parameters of the close star S2, we strongly rule out a disk optically
thick in the near infrared unless it has a relatively large inner
hole. For disks with no inner holes, we estimate that the data permit
a disk with infrared optical depth no larger than about 0.01. Such a
disk could also be responsible for the detected 3.8 $\mu$m excess in
the spectrum of S2. The constraints on the disk that we obtain here
can be reconciled with the disk parameters needed to explain the
observed X-ray flares if dust particles in the disk have sizes greater
than $\sim 30\mu$m. The destruction of small dust particles by strong
UV heating and shocks from star passages through the disk, and grain
growth during ``quiescent'' times, are mentioned as possible
mechanisms of creating the unusual grain size distribution. We
estimate the emissivity of the thin layer photo-ionized by the star in
Hydrogen Br$\gamma$ line and in the continuum recombination in the 2.2
$\mu$m band, and find that it may be detectable in the future if the
disk exists.
\keywords{accretion, accretion disks -- dust, extinction -- eclipses -- Galaxy: center --
stars: individual (S2)} 
} 
\maketitle

\section{Introduction\label{sec:intro}}

In the last decade or so, dramatic improvements in the capabilities of
the infrared instruments \citep[e.g.,][]{Ott03} produced
unprecedentedly high quality data on the distribution of stars in Sgr
A$^*$ \citep[e.g.,][]{Genzel03, Ghez03a} and in the centers of other
inactive galaxies. These high quality data prove that a super massive
black hole (SMBH) exists in the center of our Galaxy
\citep[e.g.,][]{Schoedel02,Ghez03b} and that the majority of the
bright stars in Sgr~A$^*$ cluster are early-type stars
\citep{Gezari02, Genzel03}, rising interesting questions about the
star formation history in the vicinity of \sgra.

In analogy to the disks present in LLAGN \citep[e.g.,][]{Miyoshi95,
Quataert99, Ho03}, a thin inactive (i.e. formerly accreting;
\citealp{Kolykhalov80}) disk may exist in \sgra. In addition, the disk
presence may help to explain (\citealp{Nayakshin03}; \citealp{NS03};
Nayakshin, Cuadra \& Sunyaev 2003) two major mysteries of \sgra: its
amazingly low luminosity (e.g., see \citealp{Baganoff03}, and reviews
by \citealp{Melia01} and \citealp{Narayan02}) and the recently
discovered large amplitude X-ray flares \citep{Baganoff01b}. Direct
observational detection of the disk could be possible if the disk were
massive and thus bright enough (e.g., \citealp{Falcke97} and
\citealp{Narayan02}) but a very dim inactive disk could have eluded
such a detection \citep{NCS03}.

\cite{NS03} pointed out that the {\em three dimensional} orbits of
stars such as S2 could be used to test the putative disk
hypothesis. Here we extend the work of these authors by doing a much
more through analysis of the parameter space as well as including new
physics. In particular, under the simplest assumption of an optically
thick disk, we discuss three methods to do so: (i) stellar eclipses;
(ii) re-processing of the stellar UV and visible radiation into
infrared; (iii) asymmetry of the integrated star cluster light. We
apply the first two of the methods to the star S2, whose orbit is
currently known the best out of all the close sources in \sgra\
cluster. We find (\S \ref{sec:eclipses}) that absence of eclipses for
S2 could be explained by a disk with a ``large'' inner hole, $R_{\rm
in} \simgt 0.03\arcsec$, for a rather broad range in the disk
orientations. Note that for a distance of 8 kpc to the Galactic
Center, 1\arcsec\ corresponds to $\simeq 1.2\times 10^{17}$ cm or
$\simeq 1.3\times 10^5 R_g$, where $R_g = 2 GM_{\rm BM}/c^2 \simeq
9\times 10^{11}$ cm is the gravitational radius for the $M_{\rm BH}=
3\times 10^6\msun$ black hole. The re-processing of the optical-UV
luminosity of the star in the infrared, however, creates (\S
\ref{sec:reprocessed}) a variable and very bright emission in the
standard near infrared spectral bands. This emission would add to the
total ``stellar'' flux observed from the source (see Figure
\ref{fig:lightcurvenohole} below). Considering disks with no inner
hole and all possible orientations we find that this additional
emission would have been detected by the Genzel et al. and Ghez et
al. teams. Therefore an optically thick (in $\lambda \simeq
\hbox{few}\; \mu$m) disk is ruled out for \sgra\ star cluster. We
estimate that the maximum near infrared optical depth of the disk that
would not violate observational constraints is about 0.01.  Finally,
the cluster asymmetry measurement is suggested (\S
\ref{sec:asymmetry}) as means to constrain disk size and orientation
in nearby galaxies whose nuclear regions are visible in the optical
and UV light.  In the discussion section (\S \ref{sec:discussion}) we
summarize our main results.

\section{Eclipses (and flares)}\label{sec:eclipses}

\subsection{The Method}\label{sec:method}\indent

In this paper we consider an inactive disk optically thick in all
relevant frequencies. In our study of the effects of the disk on
individual stars, we pick the star S2 \citep{Schoedel02,
Ghez03b}\footnote{\cite{Ghez03b} refer to this star as S0-2.} as the
best example. This is a very bright ($L_{\rm bol}\sim 10^5 \lsun$)
star whose orbit is constrained with better precision than that for
any other star near \sgra. The existing data on the star positions in
the last $\sim 10$ years cover \citep[ see also Figure
\ref{fig:eclipsenohole} below]{Schoedel02,Schoedel03,Ghez03b} as much
as $\sim 70$\% of its orbit. The coverage will clearly further
increase with time. In addition, the star passed mere $\sim
0.02$\arcsec\ or about $2000$ gravitational radii, $R_{\rm g}$, from
the black hole in the pericenter \citep{Schoedel02}.

\begin{figure}
\centerline{\psfig{file=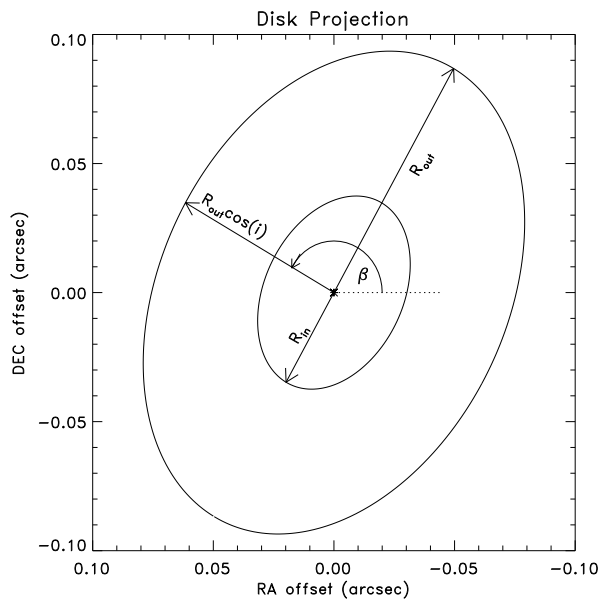,width=.45\textwidth,angle=90}}
\caption{Projection of the disk in the plane of the sky. The disk is
described by two radii, $R_{\rm in}$ and $R_{\rm out}$, and two angles: disk
inclination angle, $i$, and rotation angle, $\beta$. The latter is
defined as the angle between the West projection and the semi-minor
axis of the disk projection {\em directed to the part of the disk
closest to the observer}.}
\label{fig:cartoon}
\end{figure}

Geometrical thickness of a cold disk is very small \citep[see][]
{NCS03} so we treat the disk as a flat surface. The disk is assumed to
be in Keplerian circular rotation and is described by its inner and
outer radii: $R_{\rm in}$ and $R_{\rm out}$. An example of the
projection of such a disk is shown in Fig. \ref{fig:cartoon}. We need
{\em two} angles to describe the observational appearance of the disk:
inclination $0\leq i\leq \pi/2$, between the direction normal to the
disk and the line of sight, and rotation $0\leq \beta\leq 2\pi$,
between the West direction and the semi-minor axis of the disk as seen
in projection. (Out of the two possible semi-minor axes, we pick the
one that lies on the side of the disk closest to the observer.)

Note that the angle $\beta$ is of no importance for disks that have
azimuthal symmetry, and is thus rarely defined. For our problem,
however, this angle is important since it determines the orientation
of the disk relative to stellar orbits.  This angle definition is
somewhat different from those commonly used to define stellar orbits
\citep[see e.g.,][ \S 14.8]{Roy82}, nevertheless we feel that ours is
simpler to use: $\beta$ immediately shows which side of the disk is in
front of the plane of the sky centered on \sgra.

We use the S2 orbital parameters given by \cite{Schoedel03} and the
sign of the inclination angle measured by \cite{Ghez03b} to calculate
the 3-dimensional positions of the star as a function of time. For
each given set of the disk parameters, we determine the parts of the
star's orbit that are eclipsed by the disk (i.e. physically behind the
disk) and those that are not. A physically plausible disk should not
eclipse any of the star positions measured by
\cite{Schoedel02,Schoedel03}\footnote{\cite{Ghez03b} observed this
star in similar epochs, so using their data should yield similar
constrains.}. Note that since the Sun prevents observations of the GC
region for about half of a year, there are large portions of the
star's orbit when it simply could not be observed.

The method described in this section can be easily extended to
other stars in \sgra, once their orbital parameters are precisely
determined. S12 and S14 passed relatively close to \sgra\ so they
could also be useful in constraining the disk properties. The orbit of
S12 is indeed known quite well, except for the sign of its inclination
angle \citep{Schoedel03, Ghez03a}, so a degeneracy in the
3-dimensional positions remains.

\subsection{Sample results}\label{sec:sample}

\begin{figure}
\centerline{\psfig{file=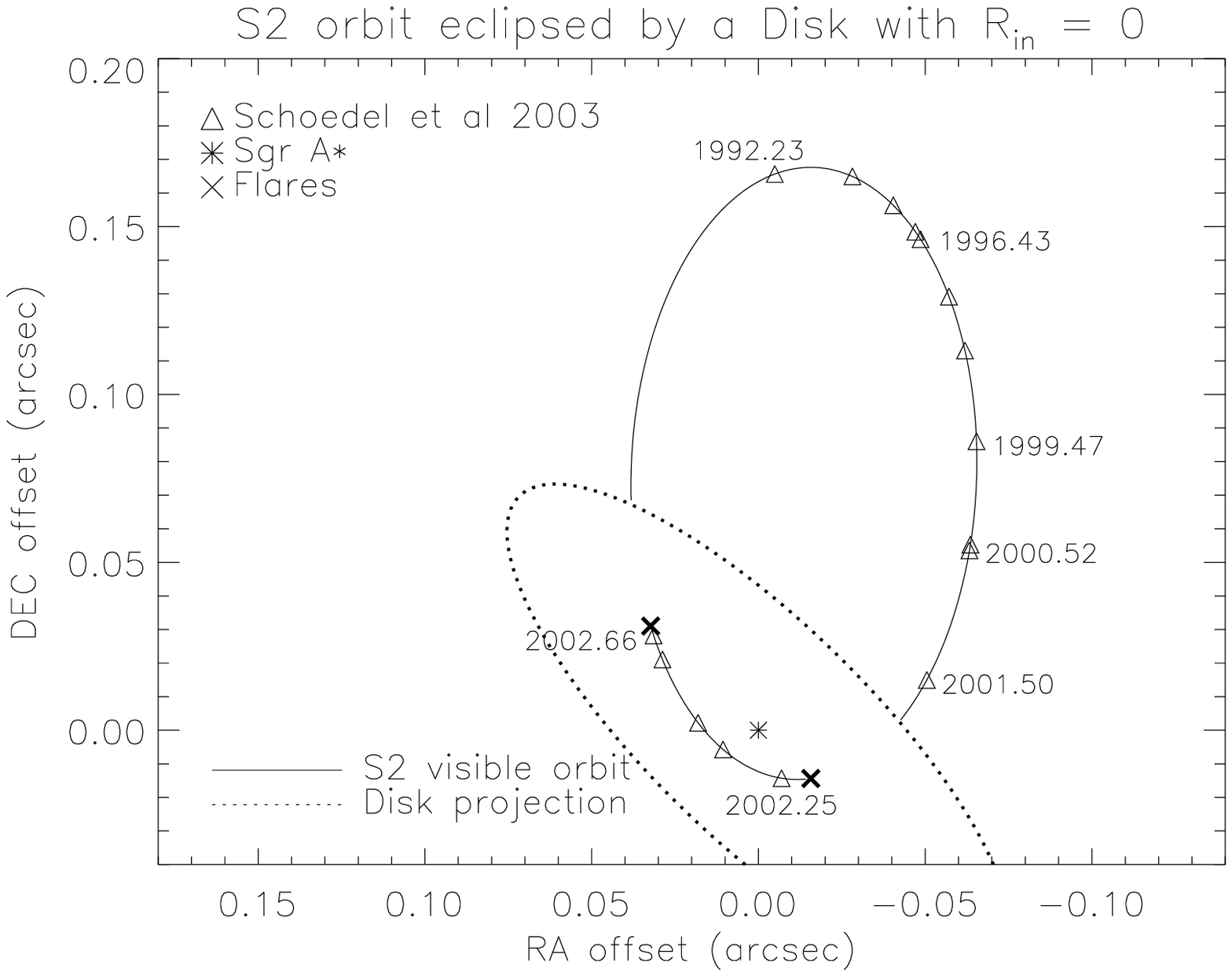,width=.45\textwidth,angle=90}}
\centerline{\psfig{file=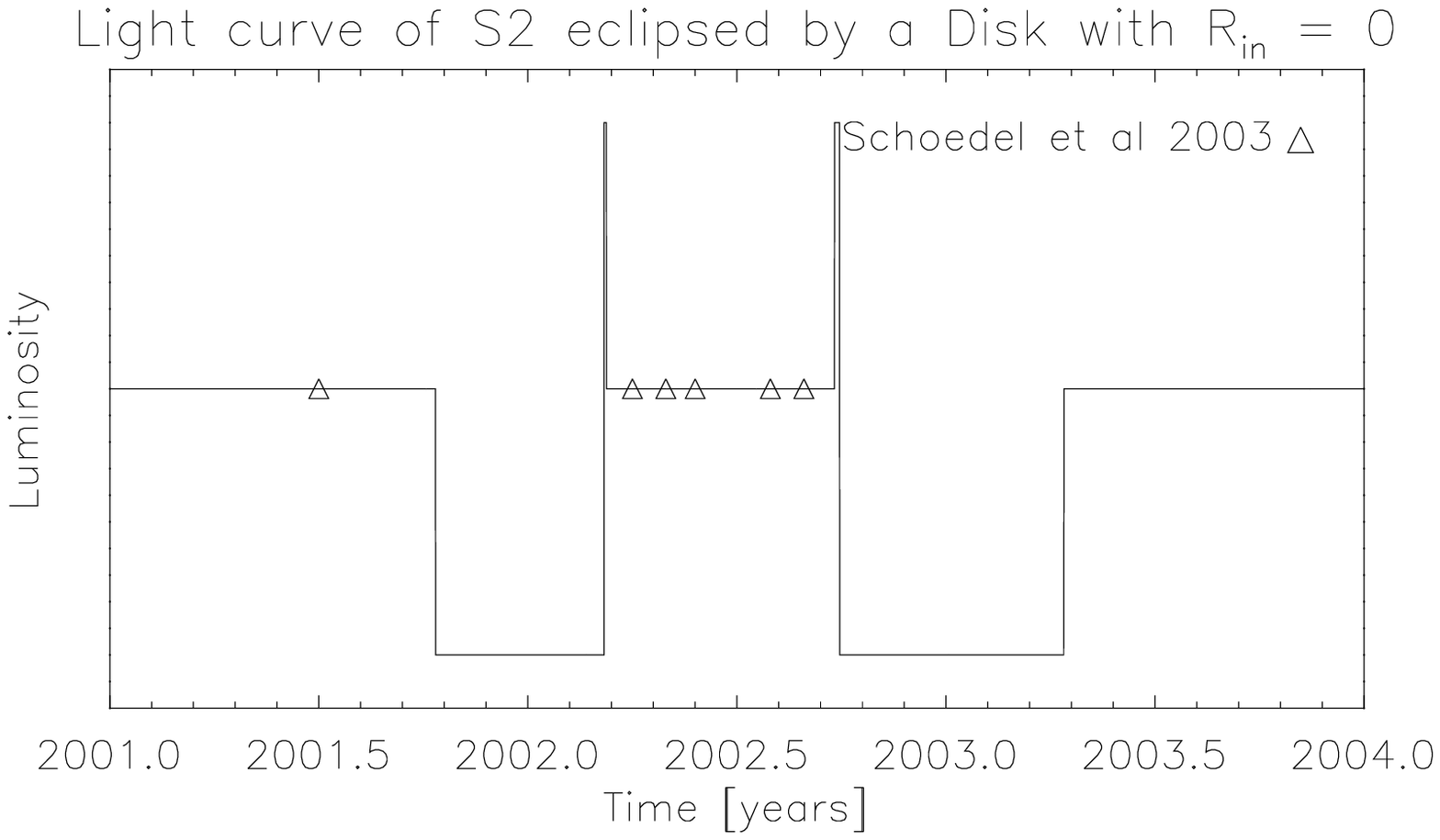,width=.45\textwidth,angle=90}}
\caption{The orbit (top panel) and the light curve (bottom) produced
by S2 passing through a disk with inner radius $R_{\rm in}=0$ and outer
one $R_{\rm out} = 0.1\arcsec$, $i$ = 71\degs\ and $\beta$ =
46\degs. Triangles show the positions of the star as given by
\cite{Schoedel03}. Respective times of the observations are shown on
the Figure for some of the measurements. The two small crosses mark
the points where the star would actually pass through the disk and
when X-ray and NIR flares are emitted. The S2 luminosity in the bottom
panel is shown schematically, in arbitrary units.}
\label{fig:eclipsenohole}
\end{figure}

To demonstrate typical results, we present two examples of disks that
are {\em not} in conflict with the observations of S2. First, we
consider a disk with no (or a very small, $R_{\rm in}\simeq 0$) inner
hole, outer radius of 0.1'', inclination $i = 71$\degs, and $\beta=
46^\circ$. In this calculation S2 was eclipsed between 2001.7 and
2002.2 (see Fig. \ref{fig:eclipsenohole}), and between 2002.75 and
2003.3. The end of the eclipse in 2002.2 coincides with a flare in
X-rays and NIR that results from the shock heating of the disk by the
star \citep{NCS03}. The NIR flare may in fact be quite strong due to
the disk re-processing of S2 radiation incident on the disk surface,
last for months, and be asymmetric (see \S \ref{sec:reprocessed}). The
same may be said about the second flare in around 2002.75, except that
for this one the eclipse begins (rather than ends) with the flare.
While fitting the observed data, this set of disk parameters is a
rather fine tuned one.  Thus such a disk (with $R_{\rm in}=0$) is
likely to be ruled out by future NIR observations of Sgr~A$^*$.

\begin{figure}
\centerline{\psfig{file=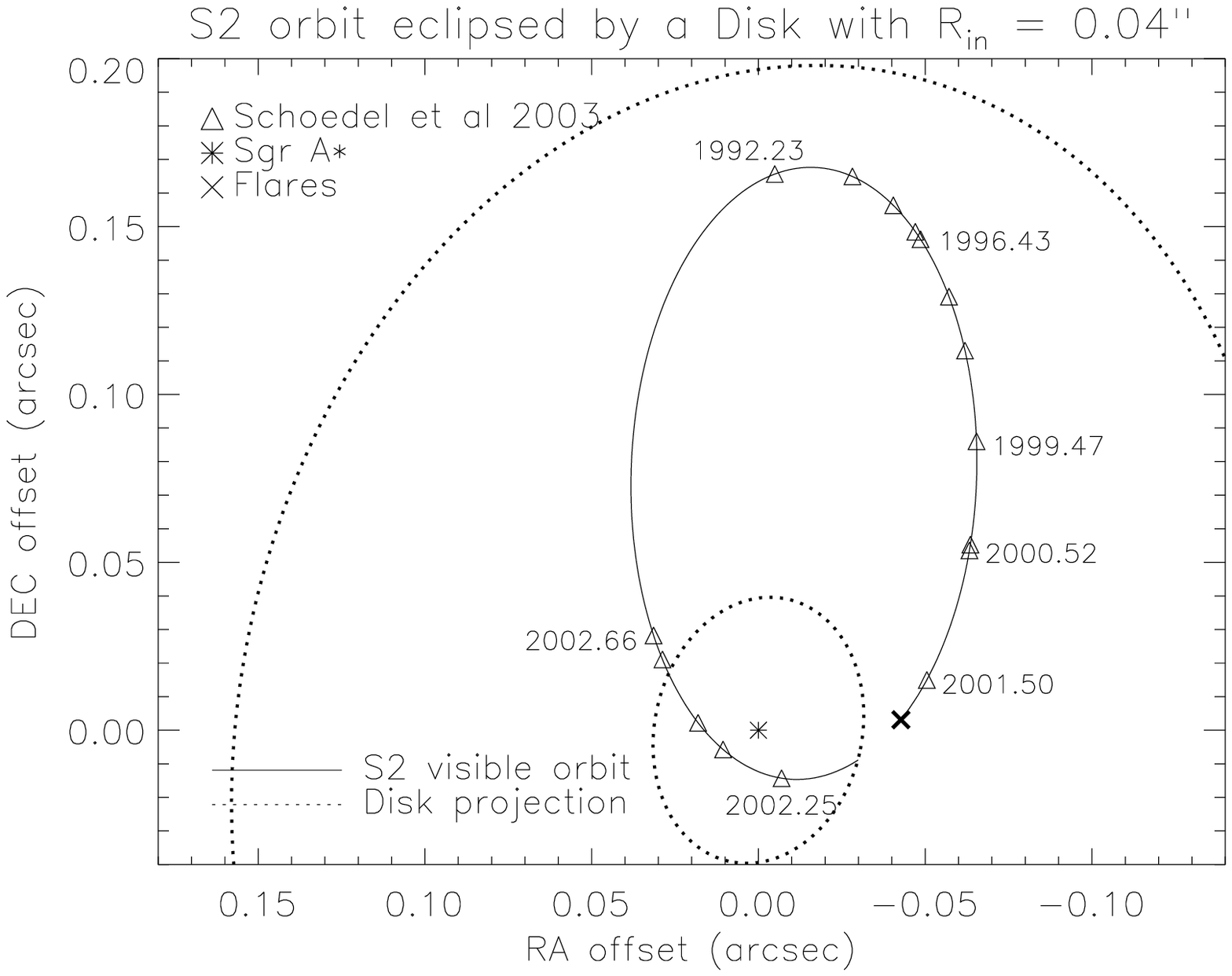,width=.45\textwidth,angle=90}}
\centerline{\psfig{file=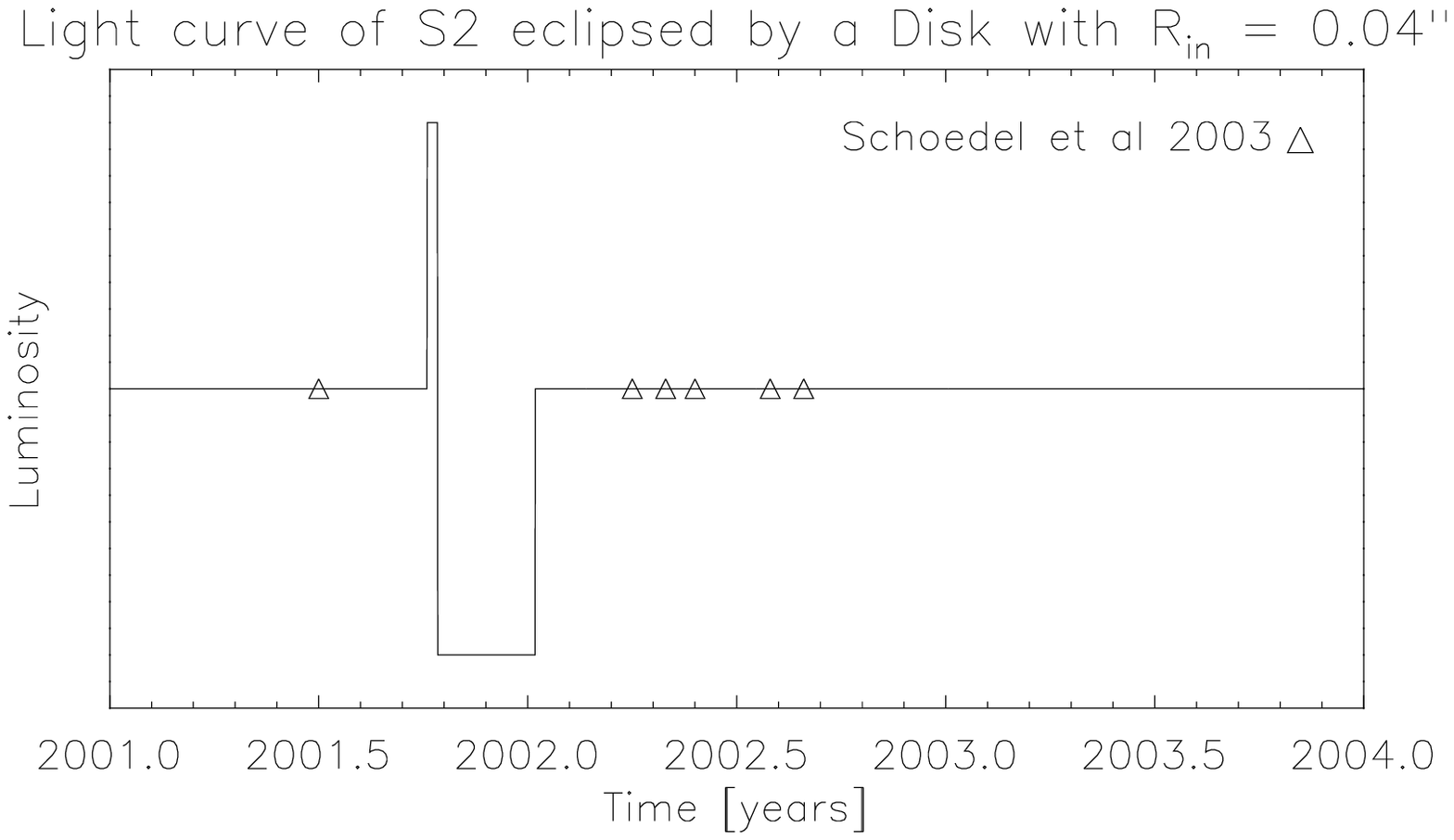,width=.45\textwidth,angle=90}}
\caption{Same as Fig. \ref{fig:eclipsenohole}, but for a disk with
inner radius of 0.04'', $R_{\rm out}= 0.2\arcsec$, $i$ = 39\degs\ and
$\beta$ = 347\degs. Note that S2 is seen through the hole in early
2002 and that there are no constrains on the disk outer radius in this
case.}
\label{fig:eclipsehole}
\end{figure}

\cite{NCS03} found that the too frequent crossings of the disk by the
close stars in the innermost region of the stellar cluster will
actually destroy the disk there.  They estimated that the inner radius
of the disk may be as large as $R_{\rm in}\sim 10^3 R_{\rm g}$,
roughly 0.01$\arcsec$. In addition, the ``accretion'' disks in LLAGN
do appear to have empty inner regions \citep{Quataert99, Ho03} with
similar values for $R_{\rm in}/R_{\rm g}$. Similarly, there are
arguments for existence of an inner hole in the standard disk
surrounding the black hole in Cyg X-1 \citep{Churazov01}. Thus it is sensible to study a disk with an inner radius $R_{\rm in}\ne
0$.  As an example, we take a much larger disk with $R_{\rm out} =
0.2\arcsec$, $R_{\rm in} = 0.04\arcsec$, and $i$ = 39\degs\ and $\beta$ =
347\degs\ (Fig. \ref{fig:eclipsehole}).  In this example the star is
eclipsed by the disk only once per orbit, not twice as in Figure
\ref{fig:eclipsenohole}. The star crossed the disk in 2001.8
(producing a flare) and appeared in the projection of the inner disk
hole in 2002.2.

An important point to take from Fig. \ref{fig:eclipsehole} is that
the outer disk radius is unconstrained in this case, i.e. it can be
arbitrarily large. This is partially due to the ``fortunate''
orientation of S2 orbit: \cite{Ghez03b} showed that out of $\sim 15$
years, the S2 orbital period, the star spent only 0.5 year behind the
black hole in the year 2002. If we take the simplest case of $i=0$,
i.e. disk coinciding with the plane of the sky, then only the
innermost 2002 points will be eclipsed if $R_{\rm in}=0$. If $R_{\rm in}$ is
greater than about $0.03\arcsec$, then the three measured positions of
S2 that were behind the plane of the sky in 2002 are observed through
the hole and the rest of the star's orbit is in front of the
disk. Therefore in this case there are no eclipses even if the disk is
infinitely large.

\subsection{Constrains on Disk Size and Orientation}\label{sec:size}

We now search the parameter space to determine the likelihood, in a
rough statistical sense, of a disk with some fixed parameters
producing no eclipses that would disagree with observations. We make a
fine grid in the parameter space ($R_{\rm in}$, $R_{\rm out}$, $\cos{i}$ and
$\beta$) for this purpose, and for each combination of these
parameters we check whether the disk eclipses any of the measured
\citep{Schoedel02,Schoedel03} star positions. If it does,
then this combination of parameters is rejected.

\begin{figure}
\centerline{\psfig{file=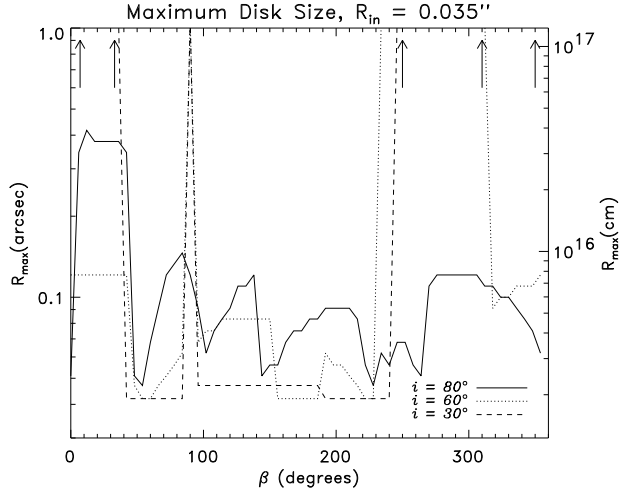,width=.45\textwidth,angle=90}}
\caption{ Maximum disk outer radius for different orientation
angles in the case of a disk with inner radius of $R_{\rm
in}=0.035''$. The maximum value of $R_{\rm out}$ is shown as a
function of the rotation angle for three different values of the
inclination angle (labelled in the Figure). The curves are rugged
because the number of the observed S2 positions is finite; have we had
a full continuous coverage of S2 orbit from 1992 to 2002, the
respective curves would become smooth functions. Vertical arrows at
the top of the plot emphasize the fact that the outer disk radius can
be arbitrarily large for the respective set of parameters. In general
much larger disks are allowed if $R_{\rm in}\simgt$ few tens of
mili-arcsecond.}
\label{fig:sizehole035}
\end{figure}

We concentrate first on a disk with $R_{\rm in} =
0.035\arcsec$. In Fig. \ref{fig:sizehole035} we show the maximum disk
size, $R_{\rm out}$, as a function of the rotation angle, $\beta$, for
three different disk inclination angles $i$. Any value of $R_{\rm
out}$ greater than the respective curve shown in
Fig. \ref{fig:sizehole035} would produce one or more observable
eclipses contradicting the data. It is seen from the Figure that
smaller values of the inclination angle generally allow larger
disks. This simply reflects the fact that the projected area of the
inner missing disk increases as $i$ decreases.

\begin{figure}
\centerline{\psfig{file=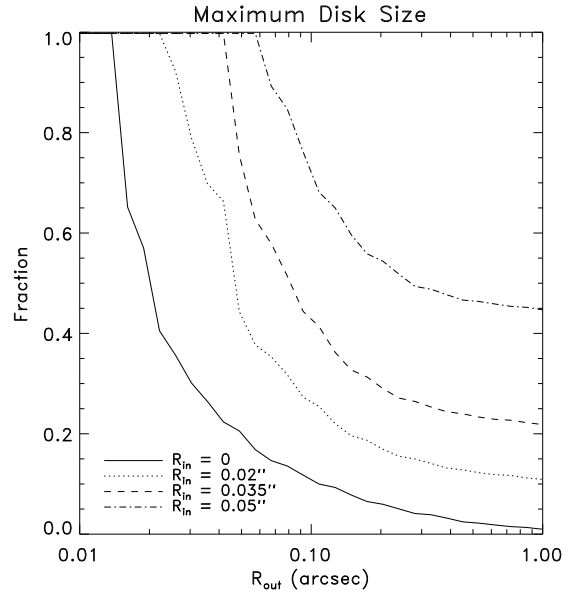,width=.45\textwidth,angle=90}}
\caption{Fraction of the parameter space that permits a disk of a
given $R_{\rm out}$. The curves are for different values of $R_{\rm in}$ as
marked in the Figure legend.}
\label{fig:fraction}
\end{figure}

If there is no physically preferred orientation of the disk, then all
the points in the $\beta$-$\cos{i}$ parameter space are
equally probable. Therefore, to give a rough statistical assessment of the
results, we may define probability $P(R_{\rm out})$ as the fraction, $F$,
of the area in this parameter space that allows the outer radius to be
larger than $R_{\rm out}$. Fig. \ref{fig:fraction} shows this fraction for
different cases (for a disk with and without an inner hole). If there
is no inner hole in the disk, then only 10\% of the disk orientations
allow outer radius $R_{\rm out}$ greater than 0.1''. When $R_{\rm in}=
0.02''$, this fraction grows to about 1/4. If the inner hole is even
larger (0.05''), only a small fraction of the orientations exclude the
optically thick disk. This result is at least partially due to the
already noted fact that S2 spends only $\sim0.5$ year behind the plane
of the sky \citep{Ghez03b}.

For the sake of the forthcoming data from the new observing season
(i.e. in 2003), we added an extra point corresponding to the predicted
S2 position on 2003.22. However the effect on the results was rather
minor -- the fraction $F$ changed by only $\sim 3\%$. Finally, since
the putative disk is very geometrically thin \citep[$H/R \sim
10^{-3}$, ][]{NCS03}, a nearly edge-on ($\cos i = 0$) disk could be
rather easily ``hidden''.

\subsection{Eclipses and flares for S2: Summary}

Using the fact that the observations of the S2 star \citep{Schoedel02,
Schoedel03, Ghez03b} showed no eclipses in the last 10 years or so we
found that an optically thick disk with no inner hole
(i.e. $R_{\rm in}=0$) is allowed for only ``small'' disks, with outer
radii $R_{\rm out}\simlt 0.1\arcsec$\ (or about $10^4 R_{\rm g} \sim 10^{16}$
cm). On the other hand, if the disk has an inner hole due to frequent
star passages or other reasons, with $R_{\rm in}\simgt 0.03\arcsec$, then
the eclipses are avoided in a large fraction of the parameter space by
disks that have relatively small inclination angle $i$. The outer disk
radius in this case may be arbitrarily large. 

It is important to point out that if the disk is optically thin, with
optical depth $\tau_K < 1$ in the near infrared $K$ band, then the
{\em eclipses are only partial} and therefore harder to observe. Due
to the results of the next section we will see that this is the most
likely situation for \sgra\ inactive disk (if there is any).

\section{Disk reprocessed emission}\label{sec:reprocessed}

\subsection{Setup}\label{sec:setup}

The putative disk is heated by the optical and UV radiation of the
member stars of Sgr~A$^*$ stellar cluster. The bulk of the absorbed
radiation will be re-emitted as the dust thermal emission in infrared
frequencies. This radiation may be observable and thus it is desirable
to determine the magnitude of the effect. In general the calculation
is by no means trivial since the disk may be optically thick in some
frequencies and yet optically thin in others, and hence a careful
treatment of radiation transfer is needed.  However, we will consider
only the case of a disk optically thick in all frequencies.

\subsubsection{Stellar spectrum}\label{sec:star}

The star S2 is identified as a massive very bright main sequence star
of stellar class between B0 and O8. In this range, the corresponding
bolometric luminosity of the star is $0.5-2 \times 10^5 \lsun$ and the
temperature is about $30,000$ K \citep{Ghez03b}.  For simplicity, we
calculate the ``model'' spectral luminosity of the star as a pure
blackbody with $T = 30,000$ K and $L_{\rm bol} = 10^5 \lsun$.
Observations of the \sgra\ star cluster in the $L'$ band
\citep{Genzel03} showed that S2 has an excess of about 0.6 magnitudes
(or 30 mJy for S2's parameters) in that band, compared with the
``normal'' colors of the surrounding stars. This excess could be due
to contamination of the star spectrum by the reprocessed emission of
the disk. What we call below the ``observed spectral luminosity'' in
$L'$ band is then the blackbody emission described above plus the
excess measured by \cite{Genzel03}.

\del{incident stellar radiation will be re-emitted as a multicolor
blackbody radiation. }

\subsubsection{The blackbody emission}\label{sec:bb}

The reprocessed disk emission is a function of time. The cooling time
of the disk is much shorter than the star's orbital period (usually by
orders of magnitude). The reprocessed emission can then be calculated
under the steady state assumption. The disk reprocessed spectrum is
thus a function of geometry, i.e. the distance between the star and
the disk, $d$. For convenience of this section, we introduce spherical
coordinate system in which the disk plane coincides with the
$\theta=\pi/2$ plane, and the star is at the $\theta=0$ axis. The
star's coordinates are thus ($r$, $\phi$, $\theta) = (d,0,0)$. (Note
that the black hole is offset from the center of these coordinates.)

We assume local black body emissivity for the disk.  In this
coordinate system the distance from a point $(r,\phi,\pi/2)$ in the
disk to the star is $\sqrt{r^2+d^2}$. We thus treat the star as a
point source and the disk as an infinitely thin plane. Clearly this
approach is inaccurate for $d\simlt R_*$, but we neglect this due to
very short duration of such a close approach. The effective
temperature of a ring with radius $r$ in disk surface is
\begin{equation}
T_{disk} (r) = \left(\frac{L_\star d}{4 \pi
\sigma}\right)^{1/4}\frac{1}{\left(r^2+d^2\right)^{3/8}},
\label{Tdisk}
\end{equation}
where $L_\star$ is the star luminosity and $\sigma$ is the
Stefan-Boltzmann constant. Assuming the black body emissivity and
integrating over $r$, we arrive at the integrated multicolor black
body disk spectrum. If the disk is inclined at angle $i$, an
additional factor of $\cos i$ should be used.

\subsubsection{Photo-ionized layer of the disk}

This simple approach (equation \ref{Tdisk}) to calculating the disk
spectrum is an approximation to the more realistic situation. At the
temperatures given by equation \ref{Tdisk}, that are usually
$T_{disk}\simlt 10^3$ K, the main agent responsible for the disk
opacity and emissivity is dust.  In reality the ionizing UV flux of
the star will create a layer of completely ionized hydrogen on the top
of the disk. The dominant role in this layer is played by the gas
rather than the dust. Within this thin layer, the recombination rate
will balance the influx of ionizing photons from the star. Assuming
temperature of order $10^4$ K for the layer, we find that its column
depth is 
\begin{equation}
N_H\sim 10^{18} n_{11}^{-1} d_{15}^{-2}\quad\hbox{cm}^{-2}\;,
\label{nh}
\end{equation}
where $d_{15}$ is distance between the star and the disk in units of
$10^{15}$ cm, and $n_{11}$ is the hydrogen nuclei density in units of
$10^{11}$ cm$^{-3}$. This column depth is orders of magnitude smaller
than that of the putative inactive disk, which was estimated by
\cite{NCS03} to be in the range $N_H\sim 10^{22}-10^{25}$ cm$^{-2}$ in
order to produce luminous enough X-ray flares (depending on the
distance from \sgra\, and taking model uncertainties into account).
The thin layer will re-emit in the optical and the UV the incident
stellar radiation. Fraction of the UV flux is emitted back out of the
disk, and a fraction is emitted towards the disk, penetrating deeper
and ionizing deeper layers. The flux re-emitted below the Lyman limit
(and below the corresponding thresholds for photo-ionization of
Helium, Oxygen and other abundant elements) can penetrate much deeper
in the disk than the original ionizing stellar photons. This radiation
will be absorbed chiefly by the dust grains ``deep'' inside the disk
and then be emitted as the blackbody calculated in equation
\ref{Tdisk}.

From this discussion it is clear that in reality a fraction of the
incident UV radiation is reflected in the optical-UV
band. Correspondingly, this fraction of the incident radiation
should not be counted in equation \ref{Tdisk}. However, for the
stellar spectrum that we assume here, i.e. the blackbody with
$T=30,000$ K, only about 20\% of the energy is emitted at frequencies
above the Lyman limit, and therefore it seems that the shielding
effect of the ionized layer should not be very large. Similarly, the
dust scattering opacity could in certain wavelengths exceed that for
the dust absorption, and then a significant fraction of the incident
stellar radiation flux could be reflected back with no change in
frequency. However we calculated (using \citealp{Draine84} optical
constants and a Mie code provided by K. Dullemond) the dust opacity
for several typical grain sizes and found that this occurs in a rather
narrow range of conditions, and hence we neglected this effect.

\begin{figure}
\centerline{\psfig{file=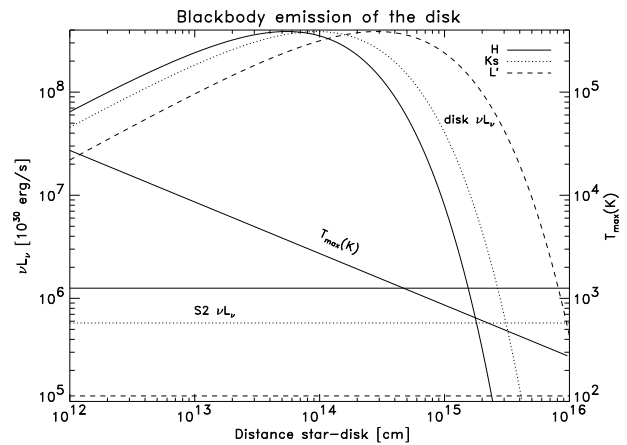,width=.45\textwidth,angle=90}}
\caption{Spectral luminosity of the disk ($\nu L_\nu$) illuminated by
star S2 in the three near infrared bands (see legend in the Figure) as
a function of the star-disk separation $d$. The disk is assumed to be
face-on to the observer ($i=0$). The model spectral luminosity (see \S
\ref{sec:star}) of S2 is also shown on the bottom of the plot. The
diagonal line shows the maximum gas temperature in the disk for the
given distance (the scale is on the right). Note that the star is
easily out shined by the disk for $d \simlt \hbox{few} 10^{15}$ cm.}
\label{fig:nuLnuvsdist}
\end{figure}

We also estimated the Brackett $\gamma$ ($\lambda=2.16 \mu$m)
line flux from this photo-ionized layer of gas. We used the number of
ionizing photons appropriate for a B0 star, a layer temperature of
10,000 K and the case B approximation \citep[see][]{Osterbrock89}. The
resulting equivalent width of this line\footnote{This is so for any
distance $d$ because the line luminosity is $\propto j_{Br \gamma} \;
\pi d^2 n_{11} N_H \propto$ const $j_{Br \gamma}$ according to
equation \ref{nh}} is $\sim$ 60 \AA, about twenty times larger than
the absorption in the Br$\gamma$ line from the star itself \citep[see
Fig. 1 in][]{Ghez03b}.  In addition, we estimated the continuum
free-free emission from the photo-ionized layer to yield $\nu
L_\nu\sim 10^{35}$ erg/sec at 2.2 $\mu$m, which is at the level of a
few tenths of S2 spectral luminosity (see Fig. \ref{fig:nuLnuvsdist}),
above the current uncertainties in the flux of S2
\citep[see][]{Ott03}. Thus, if such a strong Brackett $\gamma$ and the
free-free continuum emission from the disk were present, they should
have been detected by now.

However, the above estimates are extremely sensitive to the star's
spectrum. \cite{Ghez03b} conclude that S2 is an O8-B0 main-sequence
star. If we assume that S2 spectral type is B0.5, then the amount of
photo-ionizing photons decrease by a factor of $\sim 8$. Then both the
line and the continuum disk decrease by the same factor. In addition,
the PSF in the $K_s$ band is comparable to the size of the emitting
area of the disk. It is thus possible that a significant fraction of
the ionized layer's emission would be counted as a background
``local'' gas emission. Indeed, \citep{Ghez03b} note that the
background emission accounts for as much as 50 \% of S2's flux. Hence
to answer this question quantitatively, one needs to do a much more careful
modeling of the disk emission, i.e. convolving the latter with the PSF
instead of simply adding the two components.

Finally, due to a much faster disk rotation \citep[compared to $\sim$
220 km/s for S2;][]{Ghez03b} the Br$\gamma$ emission line should be
5-10 times broader\footnote{Here we assume that the star-disk
separation is of order the star's distance to black hole, $r_b$, in
which case a large fraction of the disk is illuminated by the star. If
instead $d\ll r_b$, then the line may be mainly shifted but not
significantly broadened.} than the line from the star, making it more
in line with the observed S2 and ``local gas'' spectrum
\citep{Ghez03b}.

\subsection{Reprocessed emission of an infinite disk}\label{sec:0hole}

In this section we study the reprocessed disk emission for disks with
no inner hole and with an infinitely large outer radius (the disk is
then a plane). In Fig. \ref{fig:nuLnuvsdist} we show the disk spectral
luminosity, i.e. $\nu L_\nu$, (integrated over the $4 \pi$ steradian
of the sky) for the frequencies corresponding to the infrared bands
$H$, $K_s$ and $L'$ (1.81, 1.38 and 0.79 $\times 10^{14}$ Hz, or 1.66,
2.18 and 3.80 $\mu$m, respectively), shown as a function of the
star-disk separation $d$. The luminosity of S2 in these three bands is
also shown for comparison with the horizontal lines on the bottom of
the Figure. On the right vertical axis of the plot, the maximum
effective temperature in the disk is shown.

Figure \ref{fig:nuLnuvsdist} shows that the contribution of the disk
can not be ignored for distances $\simlt$ few $ \times 10^{16}$ cm. S2
radiates in the Rayleigh-Jeans regime in the near infrared, and
therefore only a small fraction of the star's bolometric luminosity is
emitted at these frequencies. The disk captures a half of the star's
bolometric luminosity (mostly optical-UV) and re-processes it into
much smaller frequencies. Therefore the near infrared disk emission
can be much brighter than the star.  The distance of $10^{16}$ cm is
quite large in comparison with S2's pericenter ($\sim 3\times 10^{15}$
cm), and is of the order of S2's apocenter. Therefore the reprocessed
emission could be expected to be large for S2 for the whole of
the year 2002.

Figure \ref{fig:nuLnuvsdist} also indicates that the optically thick
disk assumption may actually break down when the star is too close to
the disk. Some of the dust species are destroyed (evaporated) when the
dust temperature is greater than several hundred Kelvin, and at $T\sim
1,500$ K the dust can be nearly completely destroyed. Therefore our
treatment is not accurate for $d\simlt 10^{15}$ cm, where the results
will be dependent on the exact disk column depth, dust properties,
etc. We overestimate the disk emission at these small distances, and
hence the maxima reached by the curves in Figure
\ref{fig:nuLnuvsdist} will be in reality smaller by factors of few to
ten. Nevertheless it is clear that the disk emission would still
dominate over the S2 emission and this fact will be sufficient for our
further analysis.

\begin{figure}
\centerline{\psfig{file=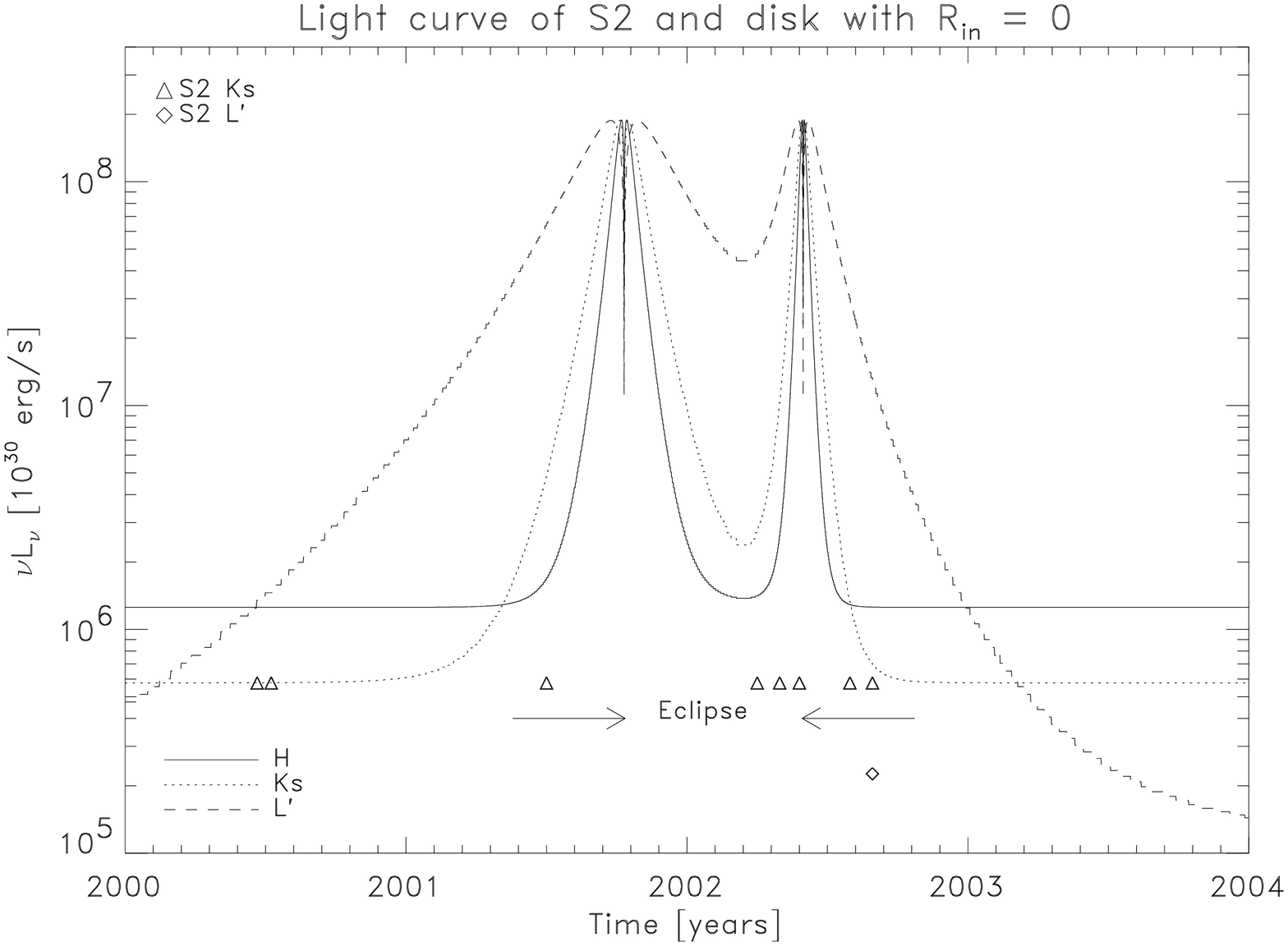,width=.45\textwidth,angle=90}}
\caption{NIR light curves of S2 and the disk reprocessed emission for
$i$ = 60\degs\ and $\beta$ = 300\degs. The triangles show the epochs
in which \cite{Schoedel02,Schoedel03} observed this star. The
corresponding ``observed'' stellar luminosity is the model spectral
luminosity in the K$_s$ band calculated as explained in \S
\ref{sec:star}. The diamond shows the model spectral luminosity of S2
plus the excess in the $L'$ band as observed (\citealp{Genzel03}; note
that there exists only one detection of S2 in $L'$ band so far).}
\label{fig:lightcurvenohole} 
\end{figure}

\begin{figure}
\centerline{\psfig{file=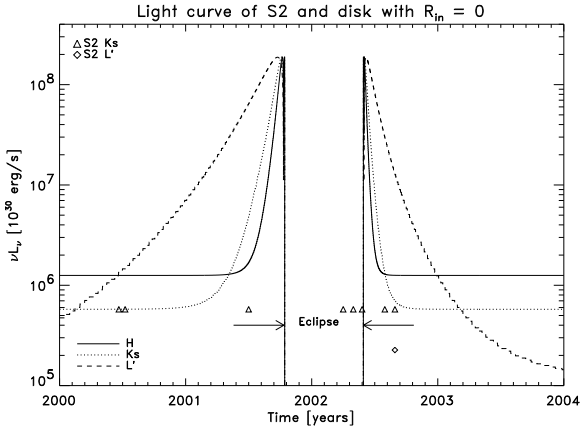,width=.45\textwidth,angle=90}}
\caption{Same as Fig. \ref{fig:lightcurvenohole}, but assuming that
both the stellar and the reprocessed emission are completely
unobservable during the eclipse.}
\label{fig:totaleclipse} 
\end{figure}

One important simplifying assumption that we make while performing
these calculations is the following. In equation \ref{Tdisk} it is
explicitly assumed that the illuminated side of the disk faces the
observer, i.e. that the star is in front of the disk. In the opposite
case the results depend on the mean optical depth of the disk,
$\tau$. Approximately, the radiation emitted from the back side of the
disk will be reduced by a factor of $\sim 1/\tau$. In what follows we
neglect this effect, i.e. assume that the front and the back side of
the disk radiate the same spectrum which is roughly correct for $\tau$
not too much larger than unity. In the following figures, however, we
point out the times when the star was behind the disk; the thermal
disk emission at these times should be remembered to be smaller than
that indicated in the figures by an amount depending on $\tau$. To
show the maximum possible effect of this complication, we simply
turned off all the NIR emission during an eclipse for one particular
calculation (see Fig. \ref{fig:totaleclipse}). Our main conclusion --
a rather unlikely presence of an {\em optically thick} disk in \sgra
-- is unchanged, and thus a better treatment of the back-side
illuminated disk is not necessary.

\subsection{The reprocessed disk emission for S2's orbit}\label{sec:rs2}

We now calculate the combined star plus disk luminosity for S2 (Figure
\ref{fig:lightcurvenohole}), first assuming that the disk is inclined
at $i=60$\degs\ and that rotation angle $\beta=300$\degs. The
triangles show the times when \cite{Schoedel02,Schoedel03} actually
observed the star. The maximum near infrared luminosity reached by the
source is the same, roughly half the star's bolometric luminosity, in
all the three frequency bands. The maxima are reached nearly
simultaneously around the time when the star physically crosses the
disk. The very sharp drops in the disk luminosity near the maxima are
simply due to the fact that the disk becomes ``too'' hot when the star
is very close to the surface of the disk (e.g., see
Fig. \ref{fig:nuLnuvsdist}). In this case the three near infrared
bands are on the Rayleigh-Jeans part of the disk blackbody curves and
the emission is therefore weak.

The part of the light curve between the two maxima in $\simeq 2001.8$
and $\simeq 2002.4$ is the time when the star is eclipsed by the disk
so that the disk emission should be actually reduced at these moments
as we explained above. In Fig. \ref{fig:totaleclipse} we show the
extreme case when the optical depth of the disk is so high that all
the emission is absorbed by the disk material during the
eclipse. Comparing the Figs. \ref{fig:lightcurvenohole} and
\ref{fig:totaleclipse}, we observe that {\em any optically thick}
infinite disk, oriented as in these Figures, is ruled out by the
existing data. There has been no changes in S2's $K_s$ band flux down
to $\sim 10-20$\% level for all 10 years of the observations (private
communication from R. Sch\"odel).

\begin{figure}
\centerline{\psfig{file=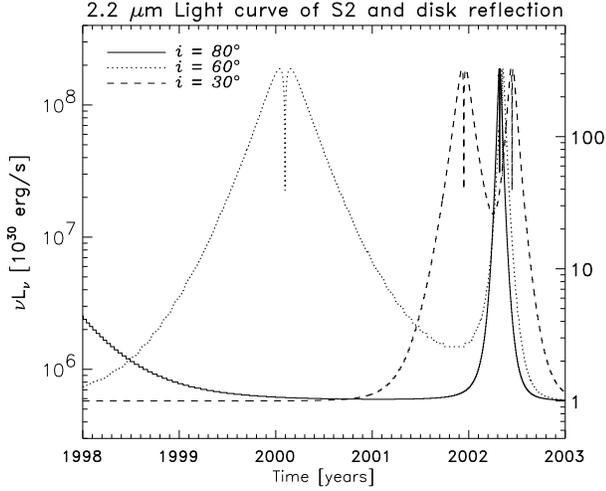,width=.45\textwidth,angle=90}}
\caption{Model spectral luminosity of S2 plus the disk reemission at
2.2 $\mu$m as a function of time for disks with $\beta$ = 80\degs\ and
different inclination angles. $R_{\rm in}=0$ for this Figure.}
\label{fig:Kslightcurves}
\end{figure}

We test the sensitivity of the result on the disk inclination angle,
$i$, in Fig. \ref{fig:Kslightcurves}, where we fix the disk rotation
angle, $\beta$ = 350\degs, but vary $i$. Three different values of $i$
(30, 60 and 80\degs) are chosen. Only the $K_s$ luminosity of the star
plus disk system is shown. The maximum near infrared luminosity
reached by the three curves is the same as in Figure
\ref{fig:lightcurvenohole}, but the times of the maxima and the width
of the curves are different. It is apparent that it is hard to escape
the tight observational constraints unless the disk oriented exactly
edge on to the observer.

\begin{figure}
\centerline{\psfig{file=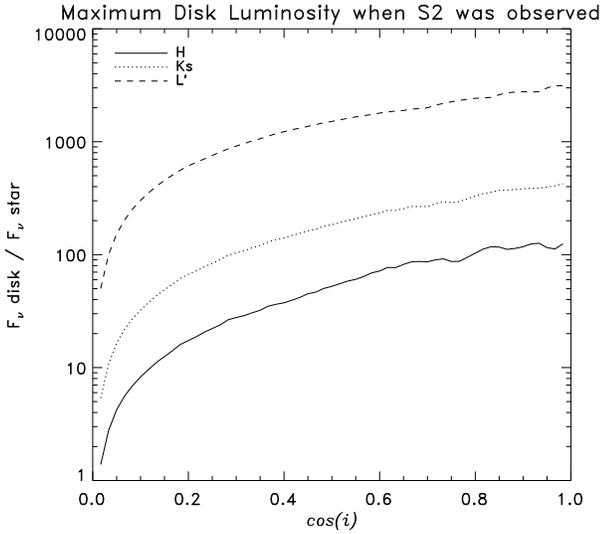,width=.45\textwidth,angle=90}}
\caption{Maximum spectral luminosity of the disk in the three NIR
bands that should been detected by \cite{Schoedel03} if an optically
thick disk with $R_{\rm in}=0$ was present in \sgra. The luminosity is
shown in units of S2 spectral luminosity as a function of the disk
inclination angle. Note that the disk is always much brighter than the
star in the infrared except for the nearly edge-on orientation.}
\label{fig:disklum}
\end{figure}

Finally, we perform a search in the parameter space in the manner
similar to that done in \S \ref{sec:size}.  In particular, we make a
fine grid in the disk orientation parameter space. For every
combination of $\cos{i}$ and $\beta$, we determine the $K_s$
luminosity of the disk at the epochs when S2 was actually observed by
\cite{Schoedel03}. We then pick the maximum of these values. The
disk $K_s$ luminosity found in this way is the maximum luminosity that
should have been observed by \cite{Schoedel03} for given
$\cos{i}$ and $\beta$. The result is displayed in Figure
\ref{fig:disklum} where we show the ratio of the disk flux to that of
the star for the three NIR frequencies, averaged over $\beta$.  Except for nearly edge-on
disks, the reprocessed emission should have been detected by
now. Since this effect has not been observed, we can rule out the
existence of an optically thick disk with no inner hole in Sgr~A$^*$.

\subsection{The case of S2 and disks with empty inner
regions}\label{sec:empty}

\del{\begin{figure}
\centerline{\psfig{file=lightcurvehole025.ps,width=.45\textwidth,angle=90}}
\caption{IR light curves of S2 and the reemission of a disk with $i$ =
60\degs\ and $\beta$ = 300\degs.}
\label{fig:lightcurvehole025}
\end{figure}
}

\begin{figure}
\centerline{\psfig{file=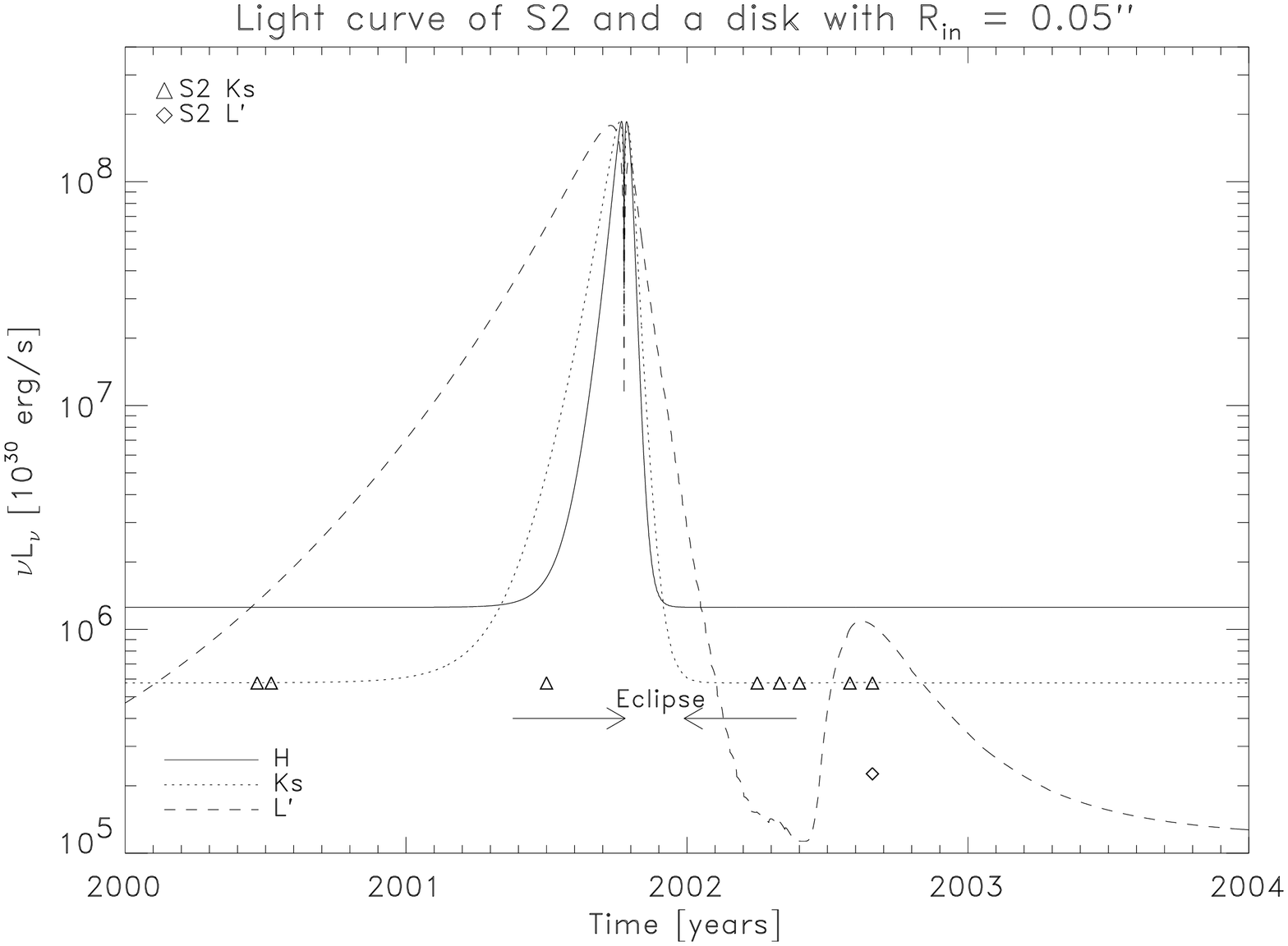,width=.45\textwidth,angle=90}}
\caption{Same as Fig. \ref{fig:lightcurvenohole} but for a disk with
an inner hole of $R_{\rm in}=0.05''$. Note the absence of the second
maxima in the $H$- and $K_s$ bands. However the $L'$ band excess is
still much larger than observed.}
\label{fig:lightcurvehole05}
\end{figure}

We now perform similar calculations but allowing the disk to have an
inner hole of a given size $R_{\rm in}\ne 0$. Figure
\ref{fig:lightcurvehole05} shows the light curves in the three
frequency bands for the disk inclined at $i=60$\degs\ and
$\beta=300$\degs, with $R_i=0.05$\arcsec. Comparing the light curves
with those shown in Fig. \ref{fig:lightcurvenohole}, the most
striking difference is the absence of the second maximum in the $H$ and
$K_s$ bands. This is due to the fact that there is now only one crossing
of the disk with the star -- in 2001.8 -- while the second crossing
shown in Fig. \ref{fig:lightcurvenohole} does not occur because
there is no inner disk for $R < R_{\rm in}$. Nevertheless the strength
of the first maximum is such that such a disk is still ruled out.

\begin{figure} 
\centerline{\psfig{file=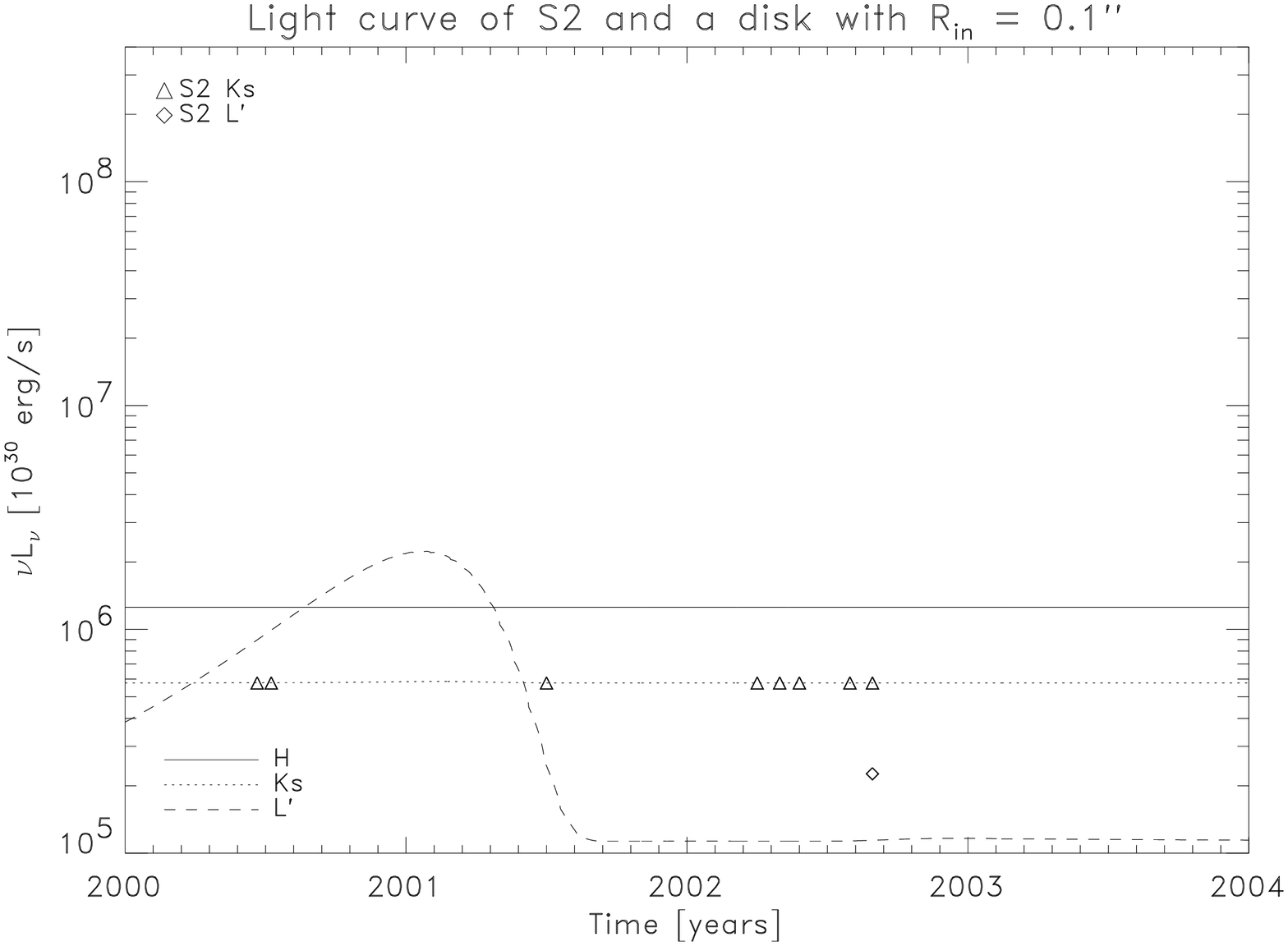,width=.45\textwidth,angle=90}}
\caption{Same as Fig. \ref{fig:lightcurvehole05} but for a larger
inner disk hole, $R_{\rm in}=0.1$\arcsec. Note that such a disk does
satisfy the observational constraints: there are no eclipses or
transient brightening in the $K_s$ band due to disk re-emission of
the star's radiation.}
\label{fig:lightcurvehole1}
\end{figure}

We then repeated the same calculation (same $i$ and $\beta$) but with
a larger inner hole radius, $R_{\rm in}=0.1$\arcsec. The result is shown
in Fig. \ref{fig:lightcurvehole1}. Since the star is relatively far
from the disk surface for all of its orbit for these particular disk
parameters, there are no eclipses or detectable increases in the $H$ or
$K_s$ bands. Further, until 2002, the S2 star has not been detected in the
$L'$ band, therefore the light curves in Fig. \ref{fig:lightcurvehole1}
appear to be consistent with the observations. In fact, we feel that
by adjusting the disk parameters it is possible to obtain the $L'$ band
spectral excess similar to that observed by \cite{Genzel03}. We
shall explore this in detail in the future papers.

\subsection{Summary: constraints due to disk reprocessed
emission}\label{Sec:sumrep}

We have shown that the reprocessed emission of the disk illuminated by
the star is a very powerful constraint on the disk presence and/or
properties. In fact, the effect is ``stronger'' than stellar eclipses
that we studied in \S \ref{sec:eclipses} \citep[see also][]{NS03}. The
luminous stars emit most of their radiation in the visible and UV
ranges. The disk re-processes this emission in the NIR band which then
appears much brighter than the star itself in the same frequency. We
have seen that the reprocessed NIR disk emission is up to a factor of
100 higher than that of the star. At the same time eclipses yield an
effect of order unity.

Analyzing the predicted NIR light curves for S2 we found that an
optically thin disk with a very small inner hole $R_{\rm in}\simeq 0$ is
ruled out. The disk emission would have been seen by now, whereas
observations do not show any variability in S2 $K$ band fluxes.  We then
tested disks with non-zero values of $R_{\rm in}$, and found that only for
rather large values of $R_{\rm in}\simgt 0.1$\arcsec\ such disks are
permitted. 

\section{Star cluster asymmetry due to stellar
eclipses}\label{sec:asymmetry}

In \S \ref{sec:eclipses} we explored the eclipses of the individual
stars by the putative optically thick disk. To observe such eclipses
one should be able to resolve individual star's orbits, which is
extremely difficult and has been made possible only for our Galactic
Center \citep[see][]{Schoedel03,Ghez03a}. At the same time, similar
eclipses of {\em unresolved} stars of the cluster should also be
occurring, and hence one may hope to detect the disk ``shadow'' on the
background emission of the nuclear cluster. In doing the calculations
below, we will be concerned with the optical-UV emission rather than
with the infrared, as we were in the previous sections. Our working
assumption is that the visible and UV flux incident on the disk is
absorbed and reprocessed into the near infrared emission (see \S
\ref{sec:reprocessed}), reflecting only a small fraction of the
radiation in the visible-UV range. Hence we treat the disk as an
optically thick absorbing surface in this section.

\begin{figure}
\centerline{\psfig{file=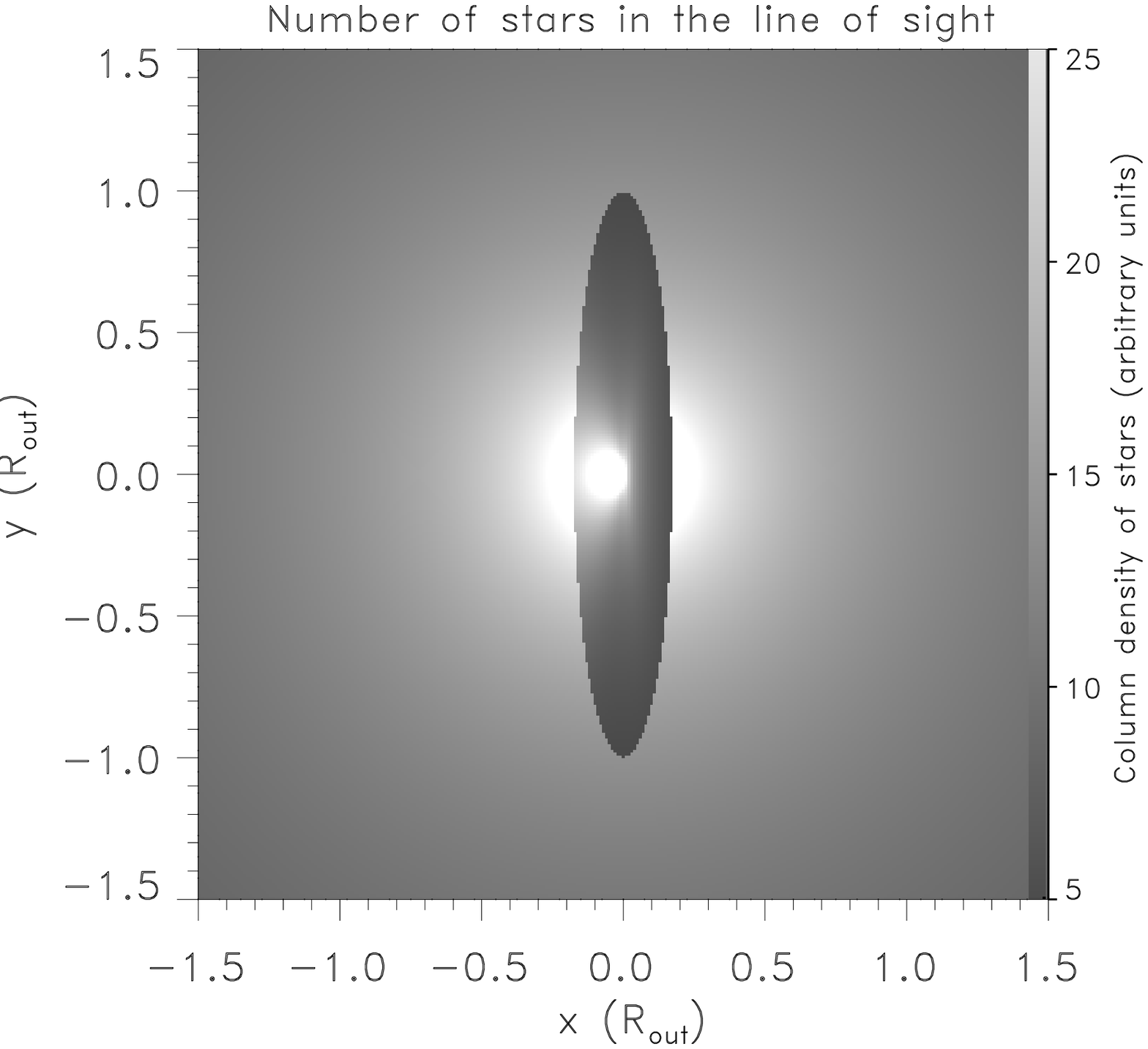,width=.45\textwidth,angle=90}}
\caption{Image of the star cluster shaded by a disk with inclination
angle $i=80$\degs. The brighter areas correspond to larger line of
sight column density of stars (the scale is shown on the right of the
Figure).}
\label{fig:shadow}
\end{figure}

We will only consider disks with no inner holes, i.e. $R_{\rm in}=0$,
since the instrument resolution (for other than \sgra\ galactic
centers) is usually worse than the actual non-zero value of $R_{\rm
in}$. 
Note that an enhanced emission from the inner hole could in principle be
detected, but {\em its asymmetry} is currently
impossible to resolve. In addition,
the star cluster is assumed to be spherically symmetric.
For this reason the angle $\beta$ is no longer of importance.  Instead we
define the $xyz$ coordinate system, with $z$ axis directed straight to
us, and $x$ and $y$ as in Fig. \ref{fig:shadow}. The $x$ axis is
positive where the disk is closer to the observer.

We also define the column density of stars along the line of sight,
$N(x,y)$, as the integral of the star density, $n(\vec{R})$, through
the line of sight from the disk to the observer, located at infinity,
\begin{equation}
N(x,y)=\int_{z'}^{\infty} n(\vec{R}) dz\;.
\label{integral}
\end{equation}
where $z'=-\infty$ if the line of sight does not intercept the disk
projection, and $z'=z_{\rm disk}$, the respective $z$-coordinate of
the disk in the opposite case.  For definitiveness, we take the
spherically symmetric density profile $n(R) \propto R^{-\alpha}$ with
$\alpha = 1.4$ for $R<R_{cusp} \sim 10''$\ given by \cite{Genzel03}
and consider the case where $R_{\rm out} \ll R_{cusp}$.

In Fig. \ref{fig:shadow} we show the resulting map of the stellar
column density for the disk inclination angle $i$ = 80\degs. The
projection of the disk shadow is clearly seen.  The boundaries of the
disk appear as sharp discontinuities in the surface brightness of the
star cluster. This effect is the strongest near the side of the disk
that is closer to the observer (positive $x$ in Figure
\ref{fig:shadow}).

Also note that the star cluster image appears anisotropic on all
scales smaller than the disk outer radius.  This fact allows us to
introduce an ``anisotropy measure'', $A$, defined as
\begin{equation}
A(r)=\frac{\sqrt{\overline{\Delta N^2}(r)}}{\overline{N}(r)}\;,
\label{a}
\end{equation}
where $\overline{N}(r)$ is the angle-averaged number count of stars at
projected distance $r$ from the star cluster center away:
\begin{equation}
\overline{N(r)}=\frac{1}{2\pi}\int_0^{2\pi} N(r,\phi) d\phi.
\label{Navrg}
\end{equation}
Here we used in the plane of the sky the common polar ($r-\phi$)
coordinates centered on the black hole.  We also defined
$\overline{N^2}(r)$ in an analogous way and then $\overline{\Delta
N^2(r)} = \overline{N^2(r)} - (\overline{N(r)})^2$.

\begin{figure}
\centerline{\psfig{file=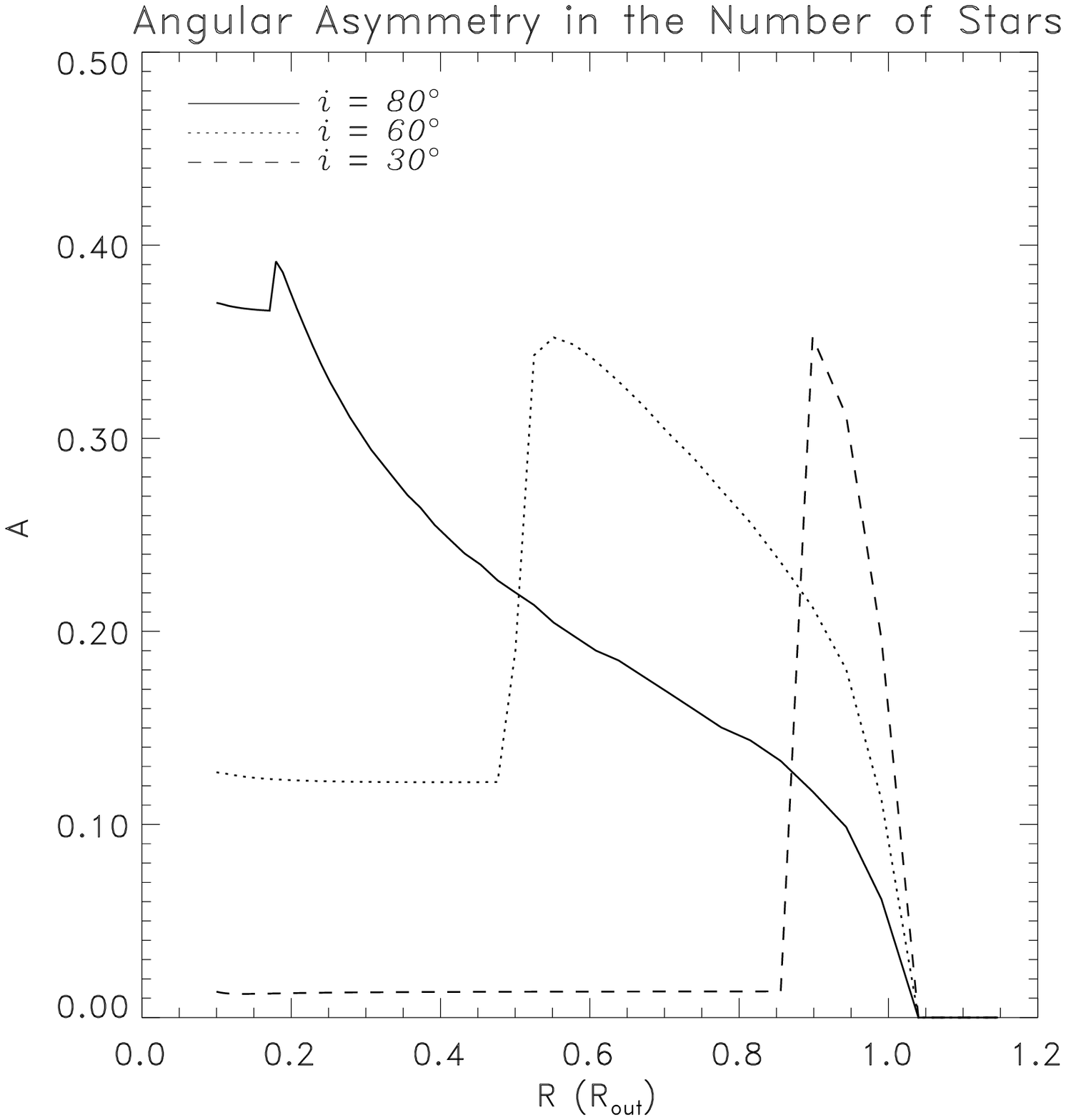,width=.45\textwidth,angle=90}}
\caption{Anisotropy measure for a ``small'' disk ($R_{\rm out} \ll
R_{cusp}$) and three values of the disk inclination angle, $i$ =
80\degs\ (solid line), 60\degs\ (dotted) and 30\degs\ (dashed).}
\label{fig:asymmetry}
\end{figure}

The anisotropy measure is independent of the disk orientation in the
plane of the sky (i.e. angle $\beta$; see Fig. \ref{fig:cartoon}).
It is also independent of the absolute luminosity of the star cluster.
Fig. \ref{fig:asymmetry} shows the anisotropy measure for three
different values of the disk inclination angle as a function of radius
$R$ in units of the disk radius $R_{\rm out}$. For small inclination
angles, $i\simlt 30$\degs, $A(R)$ is nearly zero in the innermost part
of the disk since there is little variation in $N(x,y)$ along the
circle with $R< R_{\rm out}\cos{i}$. However for $R > R_{\rm out}\cos i$, the
projected semi-minor axis of the disk, the anisotropy measure is as
high as 0.3, which is due to the high contrast between $N(R,0)$ and
$N(0,R)$. The case of the moderate inclination angle, $i=60$\degs\
shows that the anisotropy measure increases in the innermost parts of
the cluster, and there is again a maximum around
$R=R_{\rm out}\cos{i}$. This maximum appears to be the feature with which
one may attempt to identify the disk inclination angle, if radius
$R_{\rm out}$ of the disk could be inferred from independent
considerations. Finally, the case of a very highly inclined disk,
$i=80$\degs, is mostly a declining curve (after $R=R_{\rm out}\cos{i}$).

This method has the advantage that no individual stellar orbits are
needed. As the time scales for disk evolution are very long, data
collected over many years and even tens of years may be combined to
try to resolve the innermost region of the star cluster.

\section{Discussion}\label{sec:discussion}

In this paper we studied three ways to detect the disk presence in the
infrared and optical/UV frequencies, and we then applied them to the
particular case of \sgra. We found that the orbit of star S2 alone
requires the disk to be optically thin with near infrared optical
depth no larger than $\sim 0.01$. We now discuss in greater detail the
physical motivation for believing there may be a disk in \sgra, and
the implications of this paper's results for the disk hypothesis.

\subsection{A cold disk in \sgra?}\label{sec:disk}

\sgra \, is thought to be physically similar to Low Luminosity AGN
\citep{Ho99} since as long as radiative cooling is not important, the
dynamics of the accreting gas should be independent of the actual
accretion rate \cite[e.g., see review by][]{Narayan02}. \cite{Ho03}
noted that cold disks seem to be one of the established features of
LLAGN. From spectral energy distributions and from profiles of the
double-peaked emission lines, the inner radii of these disks are in
the range $\sim 10^2-10^3 R_{\rm g}$ \citep[see][] {Quataert99,
Ho03}. By analogy, such a disk can be expected to exist in the
Galactic Center.

Furthermore, \cite{NS03} suggested that star-disk crossings may be the
process that emits X-ray flares observed in \sgra \citep{Baganoff01b,
Goldwurm03}. While crossing the disk, the stars drive shocks into the
disk material; the gas is heated to temperatures of order $10^7-10^9$
K and emits X-rays. \cite{NCS03} showed that the number of star-disk
crossings per day, given by the observed distribution of stars in
\sgra \, \citep{Genzel03} is close to the observed rate of X-ray
flaring \citep{Baganoff03}; the predicted flare duration and
multi-frequency spectra are in a broad agreement with the
observations. \cite{Nayakshin03} also suggested that the disk may be
an effective cooling surface for the hot winds, and that the hot flow
is essentially frozen at large radii, preventing it from piling up at
small radii. This suggestion could be a part of the explanation for
the observed dimness of \sgra, although there are other possible
explanations \citep[see e.g., the review by][ and further references
there]{Quataert03}.

\cite{Levin03}, using data of \cite{Genzel00}, recently
concluded that most of the innermost young bright Helium-I stars (that
are thought to ``feed'' \sgra \, by producing powerful hot winds) line
up in a single plane. This result makes it very likely that these hot
young stars may have been created from a single large molecular cloud
that was compressed to high densities that led to star formation. It
is not possible that all of the gas in the molecular cloud would turn
into stars; the remainder would have to form a relatively massive
gaseous disk, similar to the bright disks of AGN. A tiny not accreted
remnant of the original disk could then still be present
\citep{Nayakshin03, NCS03}.

Finally, \cite{Genzel03} report discovery of an infrared excess in the
spectrum of S2 in 2002 (e.g., see Fig. \ref{fig:lightcurvenohole}) as
compared to other similar sources in the region. Namely, S2 appeared
to be brighter by $\sim 0.6$ magnitude in the $\lambda = 3.8\mu$m
($L'$ band) compared with what is expected from the spectra of other
bright nearby stars.  Such an excess is usually interpreted as
evidence for the dust presence around the star
\citep[e.g.,][]{Scoville76}. In the absence of the inactive (dusty)
disk, the only source of dust would have to be the hot $T \simlt 10^9
$ K flow itself \citep{Genzel03}. While this is not physically
impossible, the presence of the dust in a $T \simlt 10^9 $ K gas is
somewhat problematic as the dust can be destroyed by sputtering in
such a gas \citep[e.g.,][]{Draine79}. Our estimates show that all but
the largest interstellar grains would be destroyed by the sputtering
by the time the gas reaches radii of order S2's pericenter. We suggest
that the re-processing of the stellar radiation in a putative disk
could be an alternative way to explain the infrared excess of S2.
Whereas an optically thick disk that we explored in this paper
produces in fact a too strong an excess, an optically thin disk (with
``right'' orientation and $R_{\rm in}$ values) appears to be promising
in this regard (see Fig. \ref{fig:lightcurvehole1}).

\subsection{Constraints on the disk in \sgra}\label{sec:sgra}

\subsubsection{Optically thick disk}\label{sec:othick}

As we have shown in \S \ref{sec:eclipses}, the so far absent eclipses
of the star S2 \citep{Schoedel02, Schoedel03, Ghez03b} requires an
optically thick disk to have a relatively large inner hole $R_{\rm
in}\simgt$ few$\times 10^{-2}$ arcsecond (or equivalently few $\times
10^{15}$ cm). A hole with these dimensions is not unreasonable
\citep[e.g.,][]{NCS03}. However in \S \ref{sec:reprocessed} we showed
that the disk reprocessed emission yields an even stronger signature
than eclipses. We found that the disk with no inner hole and a very
large $R_{\rm out}$ is incompatible with the observations for any
combination of disk orientation angles. We then tested the case of a
non-zero value for $R_{\rm in}$ and found that {\em only disks with
inner holes as large as $0.1'' \sim 10^{16}\;\hbox{cm}\;\sim10^4
R_{\rm g}$ yield NIR ``echoes'' that are weak enough to escape the
observational constraints.}  As such, our results are in a complete
agreement with the previous results by \cite{Falcke97} and
\cite{Narayan02}, who also ruled out the existence of an {\em
optically thick disk with $R_{\rm in}=0$ in} \sgra.

\subsubsection{Optically thin disk}\label{sec:othin}

It is possible to make rough estimates on how optically thin the disk
should be to satisfy the observational constraints.  For this we
simply assume that the opacity of the grains is gray, in which case
the grain temperature should be equal to the effective one (equation
\ref{Tdisk}). Then the disk spectral luminosity is that calculated in
this paper but scaled down by factor of $\simeq \tau\ll 1$, where
$\tau$ is the grey disk optical depth in the NIR.  Referring now to
Fig. \ref{fig:lightcurvenohole}, we see that in the maximum the $K_s$
band disk spectral luminosity exceeds that of the star by a factor of
several hundred.  $K_s$ flux did not vary within $10-20$\% uncertainty
during the last 10 years of S2 observations (private communications
from R. Genzel and R. Sch\"odel). On the other hand the
background cluster emission accounts for as much as 50\% of the flux
in $K_s$ band \citep{Ghez03b}, and hence we estimate that the optical
depth of the disk at 2.2$\mu$m should be no larger than $\tau \sim
10^{-2}$. However, if $R_{\rm in}\ne 0$, the constrains imposed by the
2002 measured positions can be relaxed and a larger optical depth
might be consistent with the observations.

To produce X-ray flares via star-disk interactions as luminous as
observed, the mid-plane surface density of the disk was estimated by
\cite{NCS03} at around $10^{11}$ Hydrogen nuclei per cm$^{3}$
(although we note that due to simplicity of the calculations this
value is uncertain by up to a factor of 10). With this the disk
surface density, $\Sigma$, was estimated as $\Sigma\sim 1 \;
r_4^{3/2}$ g/cm$^2$, where $r_4$ is disk radius in units of $10^4$
gravitational radii, $R_{\rm g}$. For $R=0.03\arcsec\simeq 4\times
10^3 R_{\rm g}$, we have $\Sigma\simeq 0.2$ g/cm$^2$. Using the
standard interstellar grain opacity and dust-to-gas mass ratio, one
gets opacity at 2.2 $\mu$m of $\tau_{2.2} \simeq 0.6$ \citep[see
Figure 1 in][] {Voshchinnikov02}. Such a disk would violate the
constraints that we obtained in this paper. Even assuming that
\cite{NCS03} overestimated the mid-plane density by a factor of 10,
the NIR opacity still appears to be a little too large.

On the other hand, the dust grains are destroyed with each star's
passage. Smallest grains are especially vulnerable to such a
destruction. Of course when the star leaves the disk, the dust will
reform. With the gas densities as high as $10^{11}$ cm$^{-3}$, the
dust grains could grow at a rate as large as $10^{-3}$ cm/year
\citep{NCS03}.  In addition, because only the largest grains survive
the star passages, the large grains will grow preferentially.
Therefore, the combination of the repeated dust destruction and dust
growth could create larger grains than that in the interstellar
medium. Using the optical constants from \cite{Draine84} and a simple
Mie theory code to compute dust opacity (provided by K. Dullemond), we
estimated that for the standard dust-to-gas mass ratio of $0.01$ the
NIR opacity is reduced to a $\sim 1/10$th of its interstellar value at
2.2 $\mu$m if the typical grain size is $a\simgt 30\mu$m.

\subsection{Star cluster asymmetry due to disk presence}

In \S \ref{sec:asymmetry} we considered the effects of the disk on the
appearance of the integrated star cluster brightness along a given
line of sight. The disk was assumed to be optically thick and
completely absorbing at the relevant wavelength (i.e. optical or
UV). We found that the disk (obviously) imprints a shadow on the star
cluster light. We then defined the ``asymmetry measure'' parameter
$A$, defined in equation \ref{a}, which appears to be a convenient
indicator of the disk presence. In particular, as a function of radial
distance from the star cluster center, $A(r)$ has a clearly defined
shape with characteristic features that could be used to constrain the
inclination of the disk. Observational determination of $A(r)$ may
thus allow one to observationally test the disk presence in the
nuclear stellar clusters of nearby galaxies.

\section{Conclusions}\label{sec:conclusions}

In this paper we studied some of the potentially observable signatures
of a cold disk presence in our Galactic Center, and in the centers of
other galaxies. Such disks may be the remnants of previously active,
i.e. accreting, disks. This work is essentially an expansion of the
ideas presented in \cite{NS03} via (i) using the exact
and updated S2 star orbit in the study of possible eclipses of S2;
(ii) including new effects -- the disk re-processing and star cluster
anisotropy. Our present work is also complementary to that of
Nayakshin, Cuadra \& Sunyaev (2003) who studied X-ray and near infrared flares
produced when stars pass through the cold disk.

The strongest of the three effects considered in the paper turned out
to be the re-processing of the stellar visible radiation into the near
infrared bands (\S \ref{sec:reprocessed}). We found that if an {\em
optically thick disk} were present in \sgra, the reprocessed emission
of the bright star S2 would have been observed by now in all $H$,
$K_s$ and $L'$ near infrared bands. Since this contradicts to the
data, an optically thick disk would have to have a rather large inner
radius, $R_{\rm in}\simgt 0.1'' \sim 10^4 R_{\rm g} \sim 10^{16}$ cm.
 
At the same time, the observed $L'$ band excess in S2's spectrum in
the year 2002 \citep{Genzel03}, is most naturally interpreted as a
signature of the re-processing of S2 stellar radiation into this
band. We estimated that the disk invoked by \cite{NCS03} would have
the ``right'' dust opacity if the minimum size of the grains in the
disk would be about $30 \mu$m. Such a large minimum grain size may be
the result of the unusually high (by interstellar standards) density
in the disk and the too frequent stellar passages through the
disk. Therefore, {\em both the X-ray flares of \sgra \, and the
mid-infrared excess of S2 in 2002 may be the result of the
interactions of the stars with an optically thin inactive disk.}

\begin{acknowledgements}
We thank R. Genzel and R. Sch\"odel for useful comments and preprints
of their new unpublished work. We also acknowledge the expertise and
computational help of K. Dullemond in estimating the dust opacity and
the resulting disk spectrum.  Finally, we thank the referee,
W. Duschl, for comments and suggestions that improved the quality of
presentation of our paper.

\end{acknowledgements}


\bibliography{aamnem99,jcuadra} 
\bibliographystyle{aabib99}

\end{document}